\documentclass[referee,sn-mathphys,Numbered]{sn-jnl}

\usepackage{graphicx}%
\usepackage{multirow}%
\usepackage{amsmath,amssymb,amsfonts}%
\usepackage{amsthm}%
\usepackage{mathrsfs}%
\usepackage[title]{appendix}%
\usepackage{xcolor}%
\usepackage{textcomp}%
\usepackage{manyfoot}%
\usepackage{booktabs}%
\usepackage{algorithm}%
\usepackage{algorithmicx}%
\usepackage{algpseudocode}%
\usepackage{listings}%

\newcommand {\calleq}[1]{(\ref{eq:#1})}
\newcommand {\lab}[1]{\label{eq:#1}}
\newcommand {\be}[1]{\begin{equation}\lab{#1}}
\newcommand {\ee}{\end{equation}}
\newcommand {\bea}{\begin{eqnarray}}
\newcommand {\eea}{\end{eqnarray}}
\newcommand {\besys}{\left\{\begin{array}{l}}
\newcommand {\esys}{\end{array}\right.}
\newcommand {\bematr}{\left(\begin{array}}
\newcommand {\ematr}{\end{array}\right)}

\newcommand {\ind}{i}

\newcommand {\figurewidth}{0.48\textwidth}

\begin{document}

\title[Article Title]{On the role of the integrable Toda model in
  one-dimensional molecular dynamics}


\author[1]{\fnm{Giancarlo} \sur{Benettin}}\email{benettin@math.unipd.it}

\author[2]{\fnm{Giuseppe} \sur{Orsatti}}\email{gorsatti@sissa.it}

\author*[1]{\fnm{Antonio} \sur{Ponno}}\email{ponno@math.unipd.it}

\affil*[1]{\orgdiv{Dipartimento di Matematica ``Tullio Levi Civita''},
  \orgname{Universit\`a di Padova}, \orgaddress{\street{Via Trieste
      63}, \postcode{35121} \city{Padova},  \country{Italy}}}

\affil[2]{\orgdiv{SISSA}, \orgaddress{\street{Via Bonomea 265},
    \postcode{34136} \city{Trieste}, \country{Italy}}}


\abstract{We prove that the common Mie-Lennard-Jones (MLJ) molecular
  potentials, appropriately normalized via an affine transformation,
  converge, in the limit of hard-core repulsion, to the Toda
  exponential potential. Correspondingly, any Fermi-Pasta-Ulam (FPU)-like
  Hamiltonian, with MLJ-type interparticle potential, turns out to be
  $1/n$-close to the Toda integrable
  Hamiltonian, $n$ being the exponent ruling repulsion in the MLJ
  potential. This means that the dynamics of chains of particles
  interacting through typical molecular potentials, is close to
  integrable in an unexpected sense. Theoretical results are
  accompanied by a numerical illustration; numerics shows, in
  particular, that even the very standard 12--6 MLJ potential is closer
  to integrability than the FPU potentials which are more commonly used in
  the literature. }




\maketitle

\section{Introduction}\label{sec1}

The study of both the dynamical and the statistical behavior of particle
chains is a research topic essentially started by the celebrated
numerical experiment of Fermi, Pasta and Ulam (FPU) \cite{FPU},
whose aim was to study in the simplest possible examples
the time of approach to equilibrium of weakly
nonlinear systems. As is well known, on the available computation time
they did not observe any trend to equilibrium, and this is commonly
named, after the authors, the FPU problem, or paradox.

The existing literature on the subject is huge, see e.g. the
collection of papers \cite{chaos_issue,springer_fpu} or the more
recent and short reviews \cite{BCMM15,BP23}. The revival of physical
interest in such an old problem has been also stimulated, in recent
years, by the experiments on arrays of cold atoms or ions, i.e. arrays
of trapped quantum particles cooled down to extremely low
temperatures, where the lack or the slow down of thermalization is
also observed \cite{coldatoms1,coldatoms2,coldatoms3}. In the current
literature, the FPU phenomenology is referred to as
\emph{pre-thermalization} \cite{pretherm1,pretherm2}.

Nowadays, it is quite well understood that, at least on the classical
side, no paradox exists at all, the FPU problem being a manifestation
of closeness to (nonlinear) integrability, when the energy is low and
the observation time is not long enough, which means a matter of
separation of time-scales; see for example
\cite{BP11,BCP13,BPP18,CE19,GK19,GPR21,GMPR22}. More precisely, denote
by $q_\ind$ the displacement of the $i$-th particle from its reference
equilibrium position (the crystal configuration), and let $\phi(\xi)$,
$\xi=q_{\ind+1}-q_\ind$, be the interaction potential between nearby
particles. A common way (started by FPU) to express $\phi$ for small
$\xi$ is
\be{phixi}
\phi(\xi)=\frac{\xi^2}{2}+\alpha\frac{\xi^3}{3}+\beta\frac{\xi^4}{4}+
\gamma\frac{\xi^5}{5}+\cdots\ ,
\ee
with suitable
constants $\alpha$, $\beta$ and so on. Now, as pointed out
in \cite{M74}, widely developed in \cite{FFM82},
and reconsidered in recent years for example in \cite{BCP13,BPP18,CE19,GK19},
the Toda exponential potential
\be{Toda}
\mathcal{T}(\xi)=\frac{e^{\lambda\xi}-\lambda\xi-1}{\lambda^2}
\ee
for $\lambda=2\alpha$ has a contact of order three with $\phi$. But the 
Toda chain is an integrable Hamiltonian system
\cite{FST73,henon-toda,flaschka-toda,M74,FFM82}, so it is interesting to
rewrite
$$
\phi(\xi)=\mathcal{T}(\xi)+(\beta-\beta_T)\xi^4+\cdots\ ,\qquad
\beta_T=\frac16\lambda^2=\frac23\alpha^2\ ,
$$
and consider the particle chain at hand as a perturbation $O(\xi^4)$
of the integrable Toda chain \cite{BCP13,GK19,HK09}, rather than a
perturbation $O(\xi^3)$ of the harmonic chain. Since in typical
nonlocalized states (equilibrium, or excitation of some bunch of
modes) it is $q_{\ind+1}-q_\ind\sim \sqrt{\varepsilon}$, where
$\varepsilon=E/N$ is the specific energy of the system (the ratio of
the total energy $E$ to the number of particles $N$), the particle
chain \calleq{phixi} is $\varepsilon$--close to Toda, and only
$\sqrt{\varepsilon}$--close to the harmonic chain.

Having in mind Toda as the reference integrable system, the scenario
is as follows. On short terms, the dynamics of the nonlinear chain
stays close to the Toda dynamics: trajectories run on the Toda torus
corresponding to the same initial data, phases fill ergodically the
torus, while the motion transversal to tori is negligible.
Normal modes apparently interact, producing in particular the
strange energy distribution originally observed by FPU, but this
is in a sense an illusion due to the difference between linear
normal modes and conserved nonlinear Toda actions, and has nothing
to do with thermal equilibrium.  On a much larger time scale the
drift transversal to tori gets important, and diffusion in the whole
phase space, eventually leading to thermalization, does occur. Time
scales are inverse powers of $\varepsilon$, so are well separated for
small $\varepsilon$: for the standard FPU model, already on times
$O(\varepsilon^{-3/8})$ the nonuniform energy distribution observed by
FPU (the apparent paradox) starts to be evident \cite{BLP09,BP21},
while times $O(\varepsilon^{-5/2})$ look necessary to observe full
diffusion \cite{BCP13}. An intermediate time scale also exists, namely
the Lyapunov time (the inverse of the maximal Lyapunov exponent), see
\cite{BPP18,GK19,G23}; however, on such a time scale the diffusion of actions is not
affected, since chaos turns out to be tangent to tori. So, if the
observation time is sufficiently large, the FPU paradox disappears,
and in general, pre-thermalization scenarios turn out to be a matter of
interplay between observation time and dynamical time-scales of the
given physical system.

The order of contact between the pair potential \calleq{phixi} and the
Toda potential \calleq{Toda} can be increased of course by choosing
$\beta=\beta_T$ and possibly $\gamma=\gamma_T=\alpha^3/3$ and so on.
Similar exotic choices of the potential obviously lengthen the
pre-thermal scenario and the thermalization times \cite{BP11,GMMP20}.
Another possibility to approach Toda would be considering coefficients
$\beta$, $\gamma$\,...  depending on a parameter, in such a way that
in a convenient limit they all suitably converge to the corresponding
Toda values. Similar potentials might appear even more artificial and
exotic: why should physical potentials share such a bizarre property?

In this paper we address precisely this problem, and show that,
somehow unexpectedly, this latter possibility is not at all bizarre,
but is the case of a class of molecular potentials among the most used
ones, namely the Mie-Lennard-Jones potentials, and the limit in
question is that of hard-core repulsion.  This means that the above
outlined FPU scenario, with its quite long pre-thermalization times and
well separated time scales, is realistic and relevant to physics.

The Mie-Lennard-Jones (MLJ) potentials are known to model the
short-range interaction between neutral atoms or molecules
\cite{Mie,LJ31}. A possible expression is 

\begin{equation}
\lab{Mie}
\Phi_{nm}(r)=\frac{\epsilon_0}{n-m}
\left[m\left(\frac{a}{r}\right)^n-
n\left(\frac{a}{r}\right)^m\right]\ ,
\end{equation}
where $r$ denotes the inter-particle distance, $r=x_{\ind+1}-x_\ind>0$ in
dimension one, $a$ is the zero-pressure equilibrium distance, i.e. 
$\Phi_{nm}'(a)=0$, and
$\Phi_{nm}(a)=-\epsilon_0$ is the depth of the potential well,
whereas $m$ and $n>m$ are positive integers.
As is well known, the exponent $m$ rules the attractive part of the
potential due to Van der Waals charge fluctuation forces, its value
ranging from $6$ to $7$ depending on whether electromagnetic
retardation effects are taken into account \cite{DLP61}. In the present
paper we treat $m$ as a not much relevant parameter (as remarked
below, we might also allow for more general attractive tails).
On the other side, the exponent $n$, which rules
the repulsive part of potential \calleq{Mie} due to the Pauli
exclusion principle, cannot be determined from first principles, and
should be just chosen large enough to fit the experimental data on
cohesion energies \cite{K55,AM76}.
It is then natural to explore the limit $n\to\infty$ in order to see
whether an asymptotic, universal form of the interaction somehow
emerges. Such a limit amounts to model the repulsion between nearby
atoms with a hard-core barrier.

\emph{Results, in short.} In the present paper we show that, if the
potential $\Phi_{nm}$ is rescaled around the minimum (via an affine
transformation) and put in a ``normal form'' $V_{nm}$ such that the
minimum is in the origin, the second derivative
(determining the time scale) is one, and the third derivative
(determining the energy scale)\footnote{For a harmonic system, motions
  on any constant energy surface are similar to each other up to a
  trivial length rescaling, so a natural energy scale does not
  exist. Similarity is instead broken if the third derivative is
  different from zero. If $\lambda$ is the third derivative at
  equilibrium, the dominant term at small $\varepsilon$ is
  $\lambda \sqrt{\varepsilon}$; in this sense we say the third
  derivative determines the energy scale. The parameter
  $\varepsilon_0$ in MLJ is also a reference energy, but has a
  different meaning (a binding energy), is not immediately connected
  to the dynamics in the bottom of the potential well, and does not
  exist in general for potentials like \calleq{phixi} or
  \calleq{Toda}.}  fits the chosen value $\lambda$ entering
\calleq{Toda}, then it is
$V_{nm}=\mathcal{T}+O(1/n)$. Correspondingly, in the phase
space there is an affine canonical transformation depending on the
free parameter $\lambda$, which maps the Hamiltonian of the particle
chain with pairwise potential \calleq{Mie} into a new Hamiltonian
which is $1/n$-close to the integrable Toda Hamiltonian.

\vspace{2mm}\noindent {\it Remark (Toda and hard spheres:
  H\'enon's view).} \quad
Ref. \cite{henon-toda}, by M. H\'enon, is one of the three almost
simultaneous papers proving integrability of the Toda chain. It is a
short paper, in which the author does not explain how he did guess the
form of the integrals of motion. This was explained by him during a
lecture in Nice, which one of us attended. The idea
was writing down the integrals of motion for the hard-sphere gas,
which do not include potential energy and can be written as convenient
extensive combinations of velocities, and then understanding how to
compensate the lack of constancy of velocities, for the exponential
potential, by suitable mixed terms (containing products of
exponentials and velocities).  Details are not relevant to
our purposes, but it is remarkable that, in the H\'enon thought, the Toda
model was considered to be a perturbation of the hard-sphere gas (see
also \cite{flaschka-toda}, line 12). Having this in mind, it is not as
surprising that MLJ potentials, in the limit of hard-core
repulsion, get close to Toda, and a window opens to possible
generalizations.

\section{The normal form of the MLJ potentials}\label{sec2}

\subsection{Rescaling potentials}\label{subsec2.1}
Putting a potential in normal form, as outlined above, is not specific
of MLJ. Consider the class of analytic functions $f(r)$
which display a generic minimum,
namely are such that, for some $a$, 
\be{class}
f^{(1)}(a)=0\ ,\qquad f^{(2)}(a)>0\ ,\qquad f^{(3)}(a)\neq0\ ,
\ee
where $f^{(j)}$, $j\geq 1$, denotes the $j$-th derivative of $f$. 
We say that two functions $f$ and $\tilde f$ are {\it equivalent}, if they
differ just by the scale, more precisely, if they are brought one into the
other by an affine transformation:
\be{affine}
\tilde f(\xi)=A f(C\xi+D) + B \ ; \qquad A>0\,, \ C\ne0\ .
\ee
Such transformations form a group, and the class of functions
defined above gets partitioned into equivalence classes. The
transformed function
$\tilde f$ will be said to be the {\it $\lambda$--normal form} of $f$
(shortly {\it normal form}), if the minimum is
carried to the origin and the second and third derivatives are normalized: 
\be{normal}
\tilde f(0)=0\ , \qquad \tilde f^{(1)}(0)=0\ , \qquad
\tilde f^{(2)}(0)=1\ , \qquad \tilde f^{(3)}(0)=\lambda\ .
\ee

The following Lemma is easily proved: 

\vspace{4mm}\noindent
{\bf Lemma.}\ \emph{For any $f$ as above there exists an affine
transformation \calleq{affine} which gives the transformed function
$\tilde f$ the normal form \calleq{normal}. Explicitly it is 
\be{tilf}
\tilde f(\xi)=
\frac{f\left(a+\lambda f^{(2)}(a)\xi/f^{(3)}(a)\right)
  -f(a)}{\lambda^2[f^{(2)}(a)]^3/[f^{(3)}(a)]^2}
  =\frac{\xi^2}{2}+\lambda\frac{\xi^3}{6}
  +\sum_{j\geq 4}k_j\lambda^{j-2}\,\frac{\xi^j}{j!}\ ,
\ee
with
\be{qj}
k_j=\frac{[f^{(2)}(a)]^{j-3}}{[f^{(3)}(a)]^{j-2}}\,f^{(j)}(a)\ .
\ee
If in addition
$f(r)=cg(r/a)$ with $g(1)=-1$, then \calleq{tilf} and \calleq{qj} become
\be{tilf2}
\tilde f(\xi)=
\frac{g\left(1+g^{(2)}(1)\xi/g^{(3)}(1)\right)+1}
{[g^{(2)}(1)]^3/[g^{(3)}(1)]^2}
\ee
and
\be{qj2}
k_j=\frac{[g^{(2)}(1)]^{j-3}}{[g^{(3)}(1)]^{j-2}}\,g^{(j)}(1)\ .
\ee
}

\vspace{2mm}\noindent
The last statement is clearly adapted to MLJ, since
$\Phi_{nm}(r)=\varepsilon_0g_{nm}(r/a)$ with 
\be{Mie2}
g_{nm}(\rho)=\frac{1}{n-m}\left(m\rho^{-n}-n\rho^{-m}\right)\ .
\ee

\begin{proof}
The four constants $A,B,C,D$ are promptly determined so as to fit the four
requirements \calleq{normal}; expressions \calleq{tilf} and
\calleq{qj} are immediate as well. Expression \calleq{tilf2} and
\calleq{qj2} obviously follow from \calleq{tilf} and \calleq{qj}, using
$f^{(j)}(r)=ca^{-j}g^{(j)}(r/a)$.
\end{proof}

\vspace{2mm}\noindent {\it Remark (on the peculiarity of Toda potential).} \quad
Each equivalence class is characterized by the sequence
$\{k_j\}_{j\geq4}$; the class of the Toda potential \calleq{Toda} has
$k_j=1$ for any $j\geq4$. The Toda potential has another deep
property, actually used in an essential way by Dubrovin \cite{D08} to
show that the Toda chain is the unique nonlinear integrable chain
(with nearest neighbours interaction). Let us
extend the coefficient $k_4$ to the function $k_4(r)$, just by
replacing $a$ with $r$ in \calleq{qj}. For Toda it is $k_4(r)=1$
identically in $r$, and this characterizes Toda, namely imposing
$k_4(r)=1$ gives a differential equation that picks up the
exponential.

\subsection{The main result}\label{subsec2.2}
The result we shall prove is the following:

\vspace{4mm}\noindent
{\bf Proposition.} \ \emph{Consider the MLJ potential $\Phi_{nm}$,
let $V_{nm}$ be its normal
form and let $k_{nm,j}$ be the coefficients entering the series
expansion \calleq{tilf}. For any fixed $m$ and any fixed $j\ge4$ it is
\be{cnm1}
k_{nm,j}=1+O(1/n) \ .
\ee
For any fixed $m$ and any fixed neighborhood $I$ of the origin it is
\be{VnmT}
V_{nm}(\xi)=\mathcal{T}(\xi)+O(1/n)\ ,
\ee
uniformly for $\xi\in I$. 
}

\begin{proof}
It is just a computation. Using \calleq{qj2} with $g=g_{nm}$
as in \calleq{Mie2}, immediately gives 
\be{qjMie}
k_{nm,j}=\frac{(n+1)\cdots(n+j-1)-(m+1)\cdots(m+j-1)}{(n-m)(n+m+3)^{j-2}}
=\frac{n^{j-1}(1+O(1/n))}{n^{j-1}(1+O(1/n))}
\ee
and \calleq{cnm1} follows. There is no
uniformity in $j$, so this is not enough to get \calleq{VnmT}.
From \calleq{tilf2} and \calleq{Mie2}, however, we promptly obtain
\be{tilMLJ}
V_{nm}(\xi)=\frac{\left(1-\frac{\lambda\xi}{n+m+3}\right)^{-n}
  -\frac{n}{m}\left(1-\frac{\lambda\xi}{n+m+3}\right)^{-m}+\frac{n}{m}-1}
{\lambda^2\frac{n(n-m)}{(n+m+3)^2}}\ .
\ee
The denominator is clearly $\lambda^2+O(1/n)$. The first term at the
numerator, profiting of the very definition of the Euler number, gives
$e^{\lambda\xi}(1+O(1/n))$, 
while the remaining part of the numerator gives $-\lambda\xi-1+O(\xi/n)$.
The conclusion follows.
\end{proof}

\vspace{2mm}\noindent{\it Remark (on the tail of MLJ potentials).}\quad
By following the proof, it clearly appears that the first term inside
the square bracket entering the MLJ potential \calleq{Mie} produces,
in the limit $n\to\infty$, the exponential $e^{\lambda\xi}$ entering Toda
potential, while the second term provides the subtraction
$-\lambda\xi-1$. The power form of the former looks indeed essential to
work out the exponential, while the detail of the latter looks
not as relevant, and the subtraction $-\lambda\xi-1$ is generally
expected, since the normal form (for each $n$ and in the limit) must
satisfy \calleq{normal}. This means that the precise expression of the attractive
tail in MLJ is irrelevant, and in fact, it is a trivial matter to
check that the power $(r/a)^{-m}$ can be replaced by any function
$\varphi_m(r/a)$, independent of $n$, provided $\Phi_{nm}$ has a critical
point in $a$; for this it is enough $\varphi(1)=1$, $\varphi^{(1)}(1)=-m$
(the critical point is automatically a minimum for large $n$).

\vspace{2mm}\noindent{\it Remark (on the limit at constant $m/n$).}\quad
As a curiosity, we can study the normal form $V_{nm}$ of the MLJ potential, as
$n\to+\infty$, not at fixed $m$, but at fixed ratio
$\delta=m/n<1$. It is not difficult to see that 
\be{tilMielim2}
V_{n,\delta  n}(\xi)\ \longrightarrow\ \frac{(1+\delta)^2}
{\lambda^2\delta(1-\delta)}
\left[\delta e^{\frac{\lambda\xi}{1+\delta}}-e^{\frac{\delta\lambda\xi}{1+\delta}}
+1-\delta\right] \ .
\ee
An interesting choice is $\delta=1/2$, which gives
\be{Morse}
V_{n,n/2}(\xi)\ \longrightarrow\ \frac{9}{2}
\left(e^{\lambda\xi/3}-1\right)^2\ ,
\ee
i.e. the Morse potential. For $\delta\to0$, as is not surprising, the
Toda potential is recovered.

\subsection{Canonical completion of the normalization}\label{subsec2.3}

The above normalization involves only the coordinates of the particles,
but it naturally extends to momenta, also in the
limit $n\to\infty$, so as to have a
canonical transformation. The Hamiltonian of a particle chain with
nearest neighbours potential \calleq{Mie} is 
\be{molH}
H(x,p)=\sum_{\ind=0}^{N-1}\left[\frac{p_\ind^2}{2m}
  +\Phi_{nm}(x_{\ind+1}-x_\ind)\right]\ ;
\ee
to fix the ideas let us think of fixed ends, i.e. $x_0=0$
and $x_N=L$, with $L=Na$ so as to have zero pressure, and correspondingly
$p_0=p_N=0$. 

The canonical transformation can be divided into two steps. First,  
a translation followed by a rescaling of coordinates $x_\ind,p_\ind$
and time variable $t$, namely
\begin{eqnarray*}
&& x_\ind=a(\ind+Q_\ind)\ ,\qquad p_\ind=a
\sqrt{m\Phi^{(2)}(a)}\ P_\ind\ , \\
&& t=\sqrt{\frac{m}{\Phi^{(2)}(a)}}\,\tau\ , \qquad
H=a^2\Phi^{(2)}(a)\,K(Q,P)+N\Phi_{nm}(a)\ ,
\end{eqnarray*}
which is canonical with valence $a^2\sqrt{m\Phi^{(2)}(a)}$; the new
boundary conditions are $Q_0=Q_N=0$, $P_0=P_N=0$. The new Hamiltonian
reads
\[
K(Q,P)=\sum_{\ind=0}^{N-1}\left[\frac{P_\ind^2}{2}
  +\frac{\Phi_{nm}(a(1+Q_{\ind+1}-Q_\ind))-\Phi_{nm}(a)}
  {a^2\Phi_{nm}^{(2)}(a)}\right]\ .
\]
The second step consists in rescaling coordinates and momenta
by a further factor $w=\lambda a^{-1}\Phi_{nm}^{(2)}(a)/\Phi_{nm}^{(3)}(a)$,
namely
\[
Q_\ind= w\zeta_\ind\ ;\qquad P_\ind=w\eta_\ind\ ;\qquad
K=w^2\mathcal{H}(\zeta,\eta);
\]
this is canonical with valence
$w^2$, and the new Hamiltonian $\mathcal{H}$ is
\[
\mathcal{H}(\zeta,\eta)=\sum_{\ind=0}^{N-1}\left[\frac{\eta_\ind^2}{2}+
V_{nm}(\zeta_{\ind+1}-\zeta_\ind)\right]=
\sum_{\ind=0}^{N-1}\left[\frac{\eta_\ind^2}{2}
  +\mathcal{T}(\zeta_{\ind+1}-\zeta_\ind)\right]
+O(1/n)\ .
\]
Use has been made of \calleq{cnm1} and \calleq{tilMLJ}.
This proves that the Hamiltonian \calleq{molH} is, up to a canonical
normalization, $1/n$-close to the integrable Toda one.
Starting with a generalized MLJ potential, with a tail
$\varphi_m(r/a)$ as discussed above, leads to the same conclusion.  

\begin{figure}
\centering
\includegraphics[width=\figurewidth]{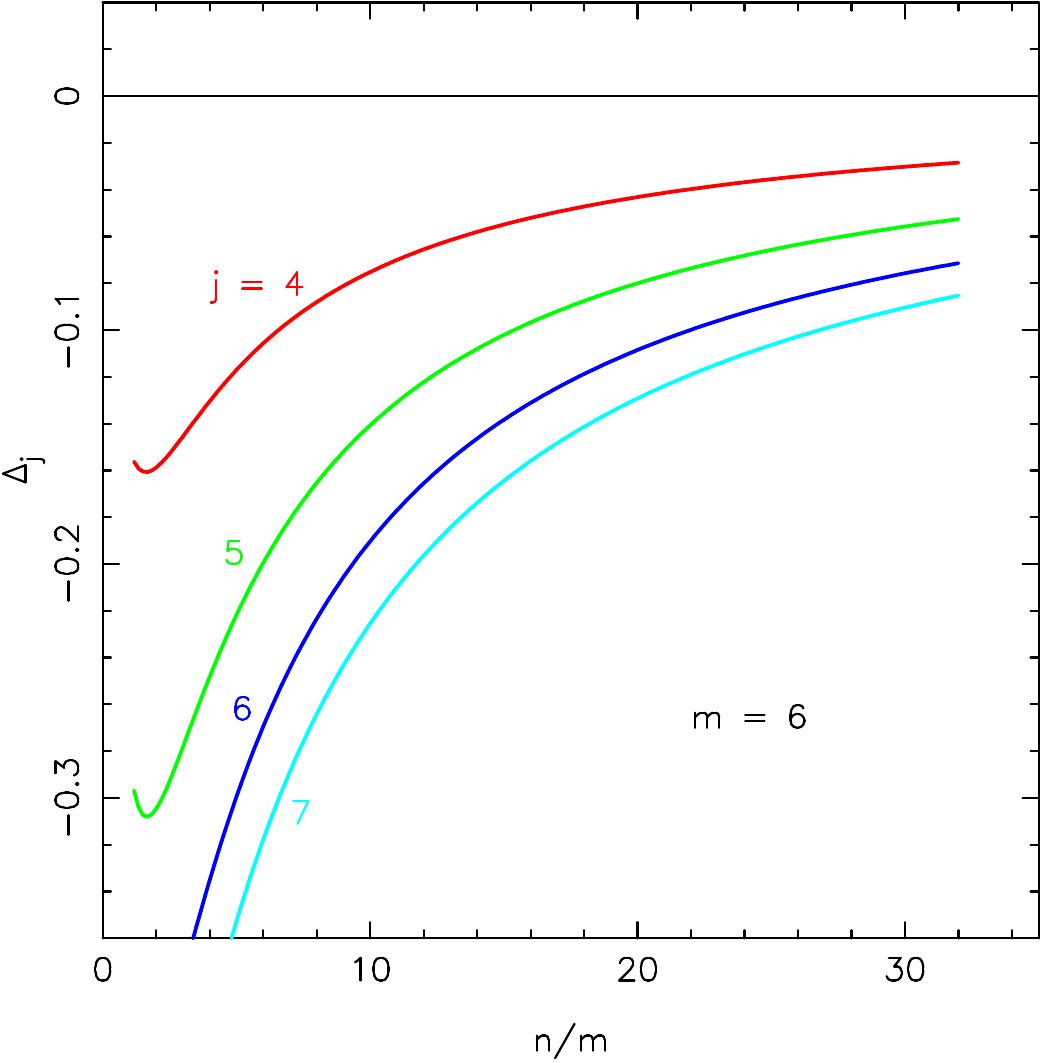}
\caption{The differences $\Delta_j=k_{nm,j}-1$ vs. $n/m$, for fixed
  $m=6$; $j=4,\ldots,7$.}
\label{f1}
\end{figure}

\begin{figure}
  \centerline{\includegraphics[width=\figurewidth]{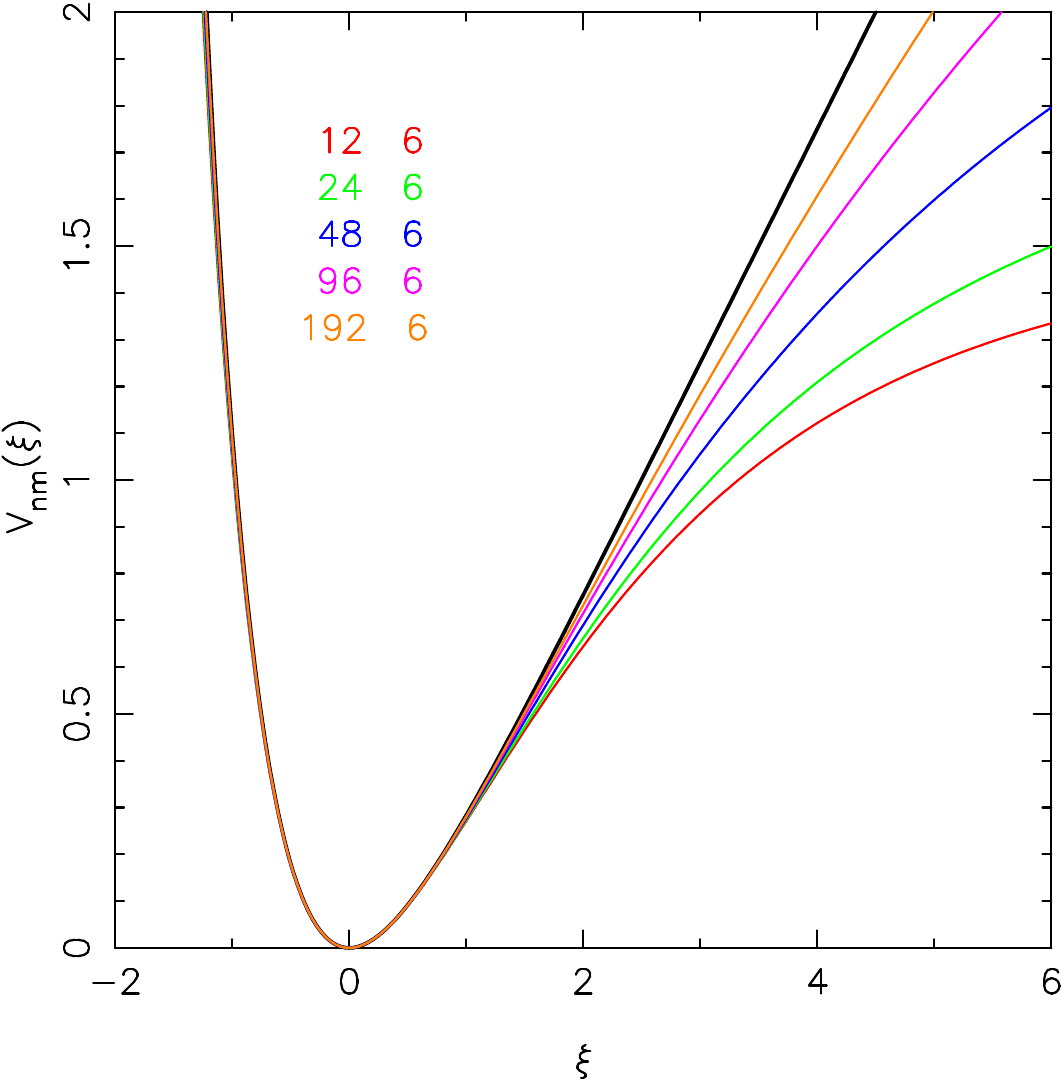}\hspace{1mm}
    \includegraphics[width=\figurewidth]{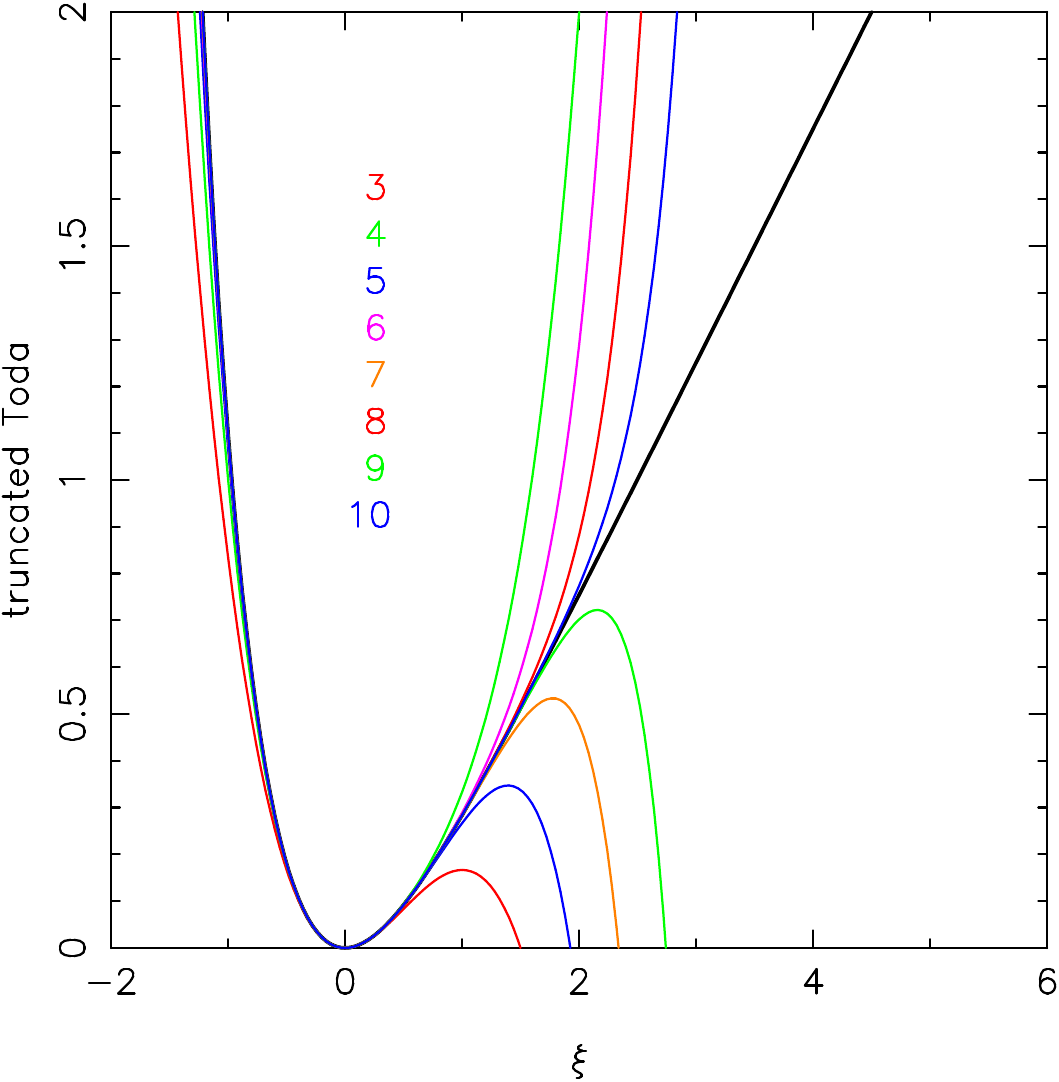}   }
  \caption{Left panel: Toda potential $\mathcal{T}$ (black) and
    normalized MLJ potential $V_{nm}$, for fixed $m=6$ and
    $n=12,\,24,\,48,\,96,\,192$ (bottom to top, colored curves);
    $\lambda=-2$. Right panel: Toda potential and its Taylor
    truncations $\mathcal{T}_j$, $j=3,\ldots,10$.  }
\label{f2}
\end{figure}

\begin{figure}
  \centerline{\includegraphics[width=\figurewidth]{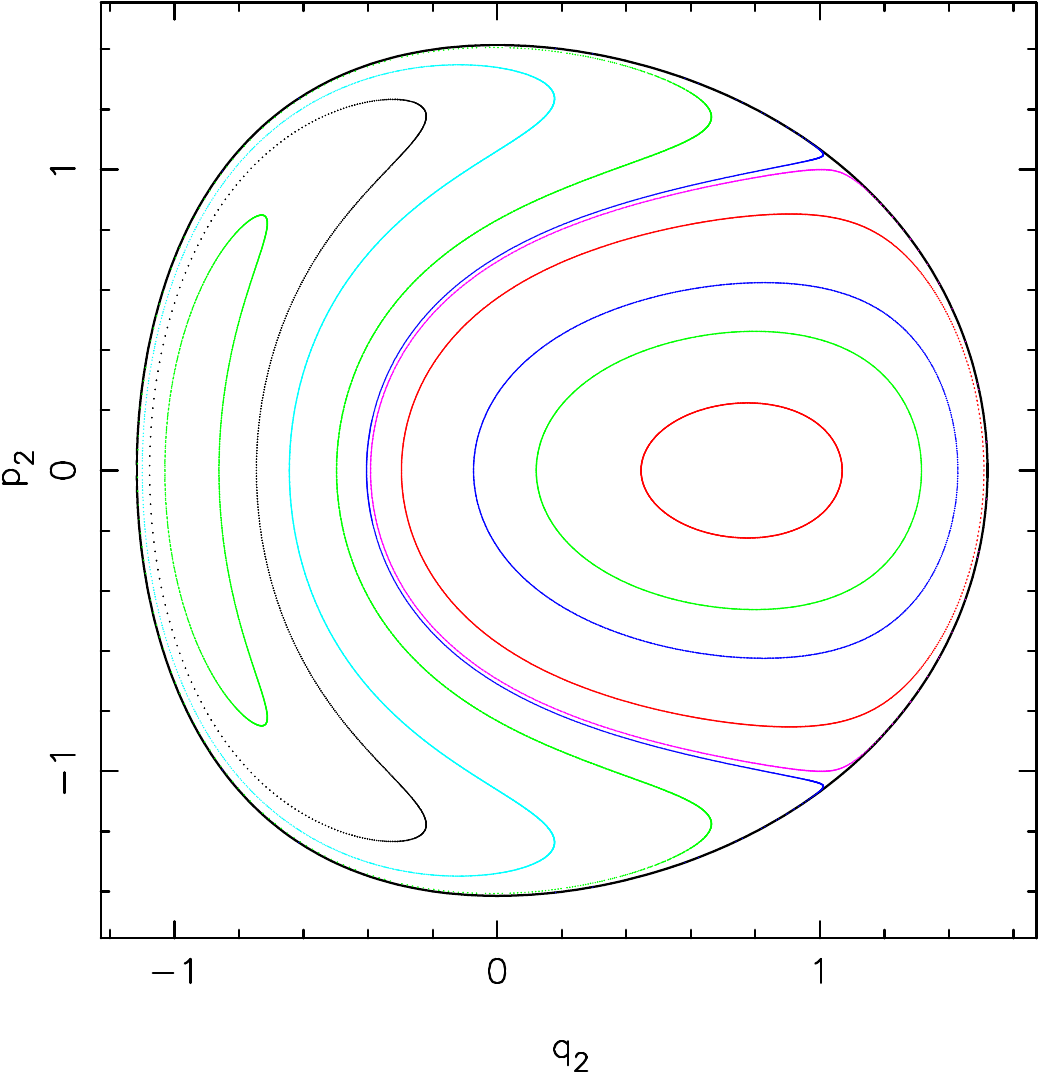}\hspace{1mm}
    \includegraphics[width=\figurewidth]{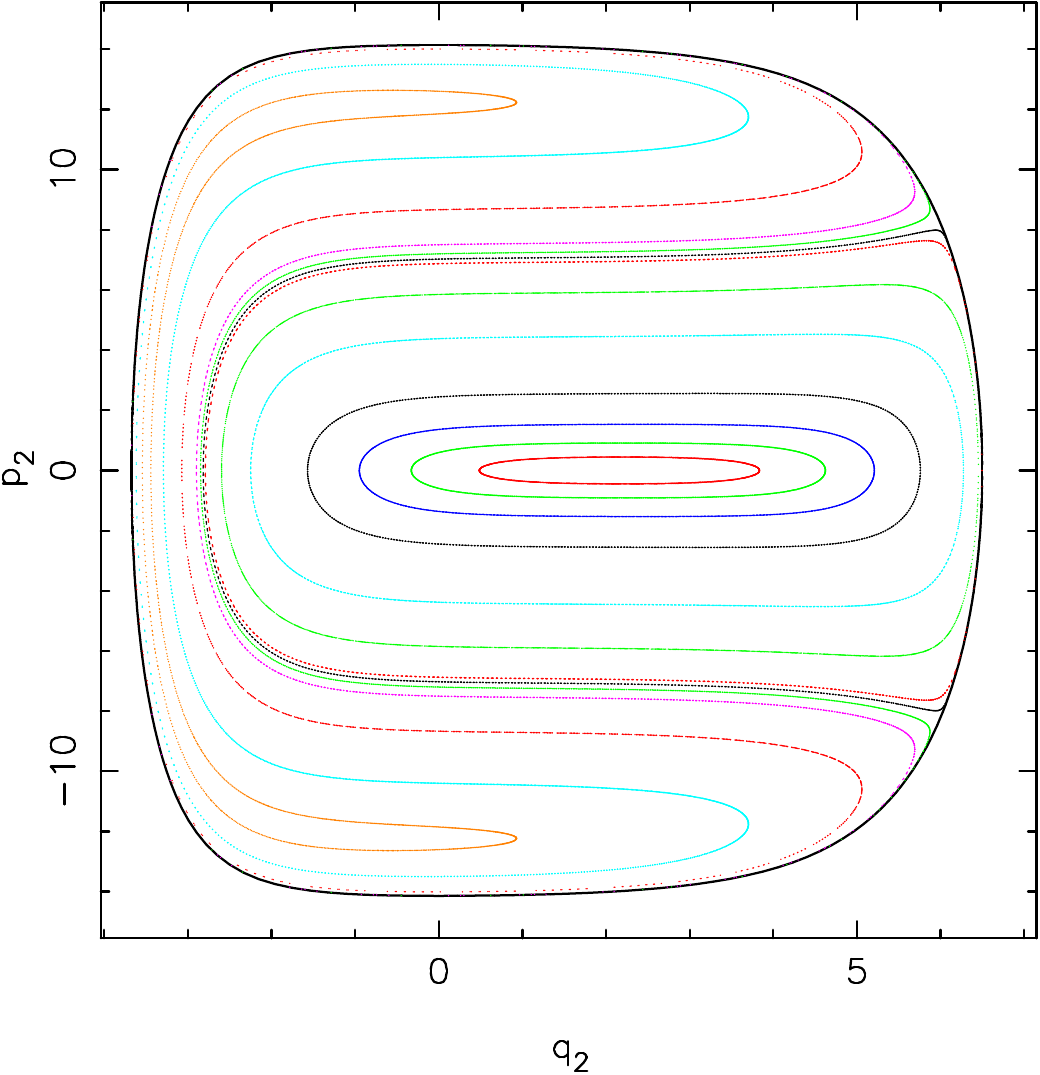}   }
  \caption{The Poincar\'e section of the model of three particles on a
    ring, with Toda potential. Left panel $E=1$, right panel $E=100$.}
\label{f3}
\end{figure}

\begin{figure}
  \centerline{\includegraphics[width=\figurewidth]{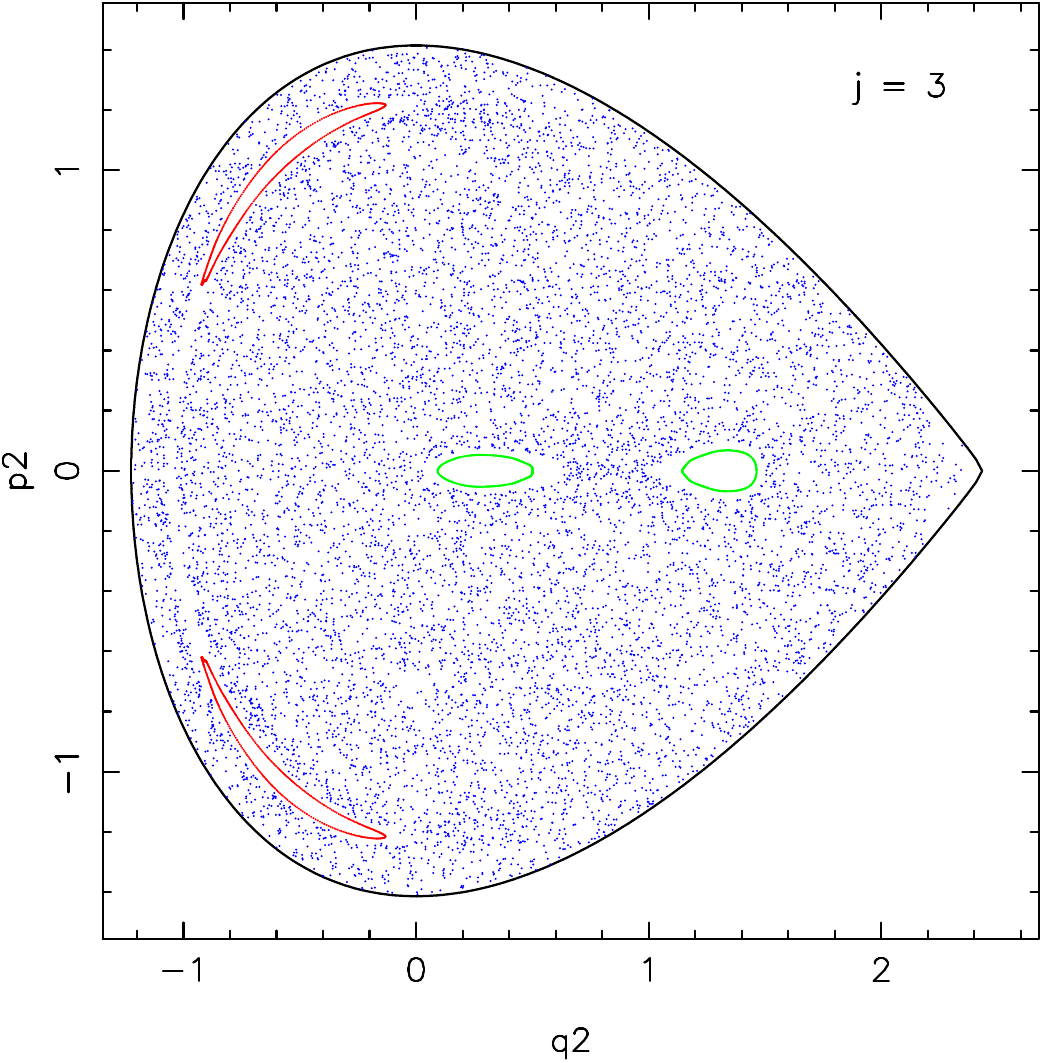}\hspace{1mm}
    \includegraphics[width=\figurewidth]{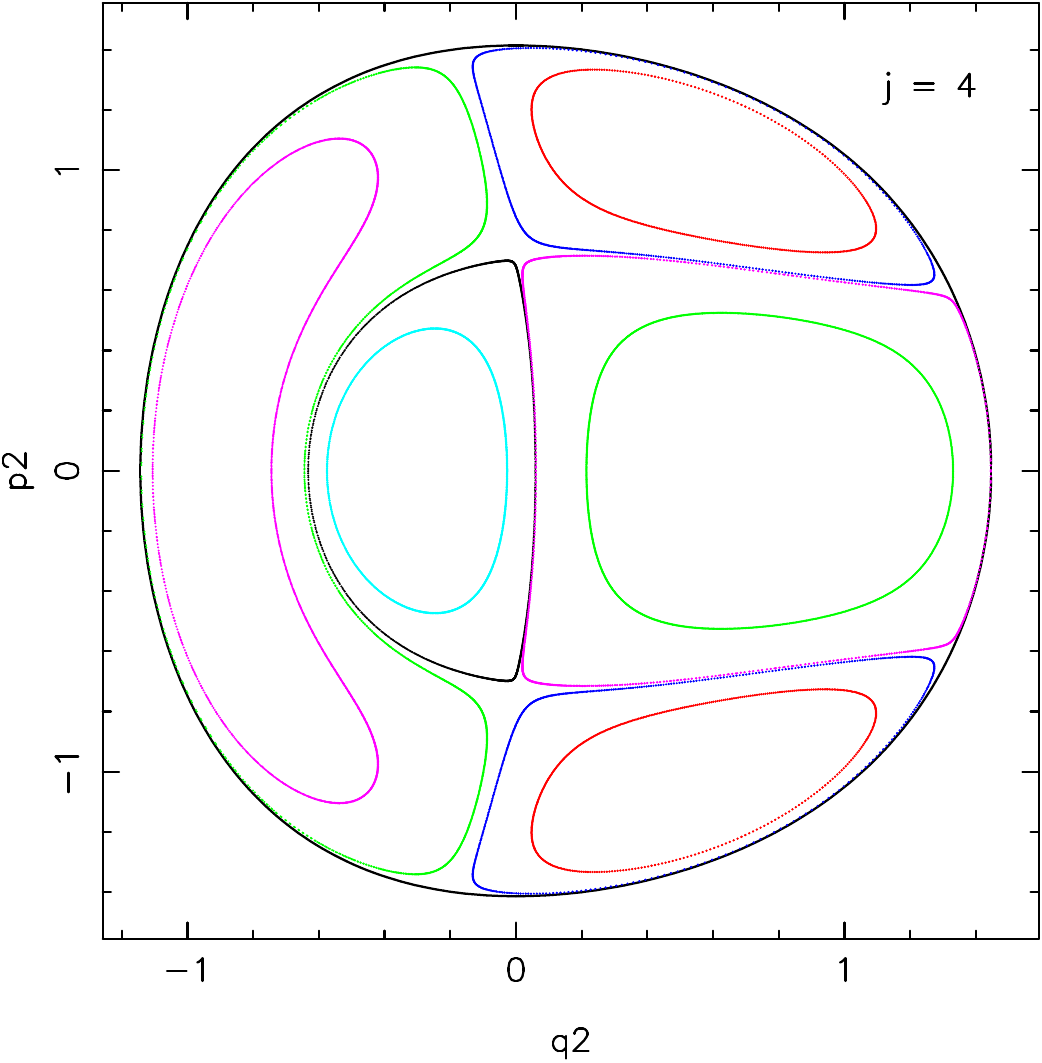}   }
\vspace{3mm}
  \centerline{\includegraphics[width=\figurewidth]{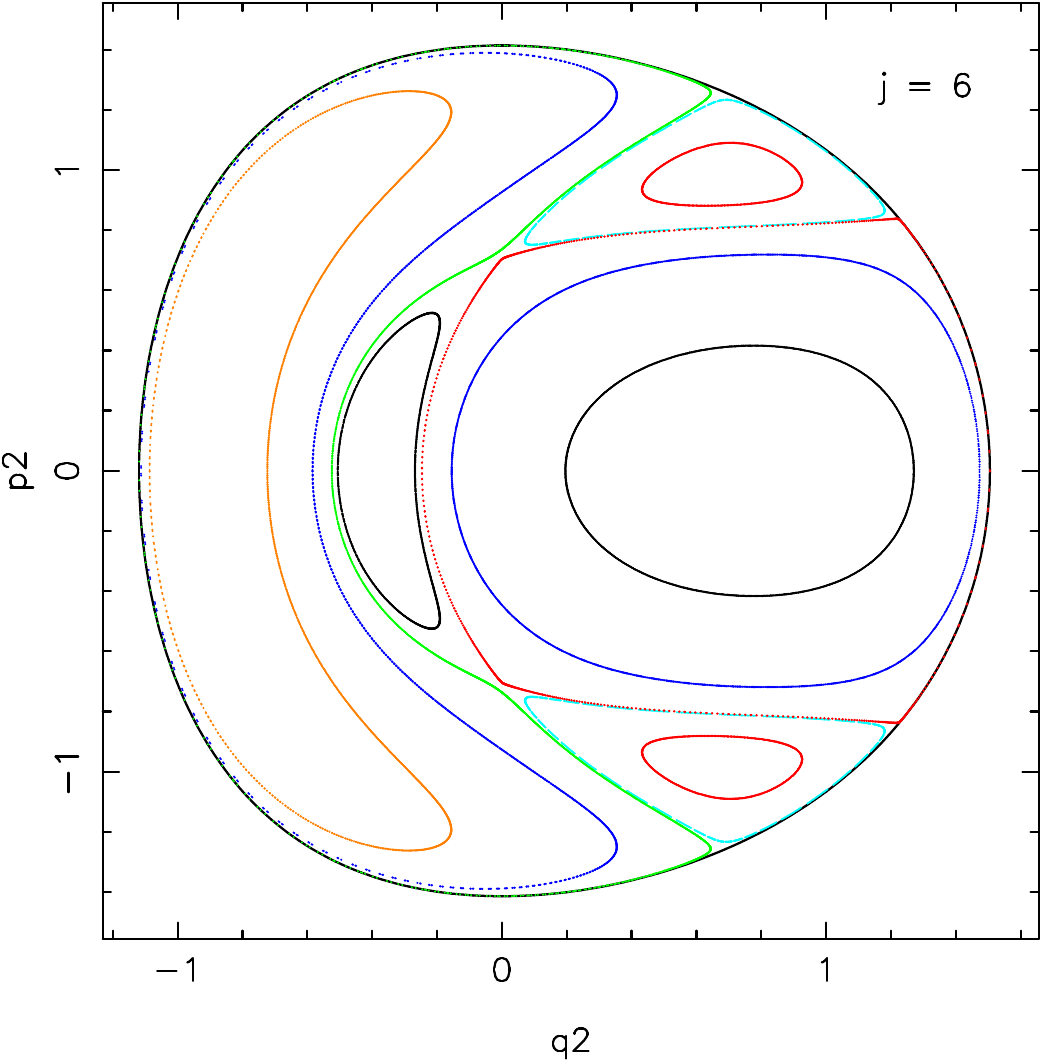}\hspace{1mm}
    \includegraphics[width=\figurewidth]{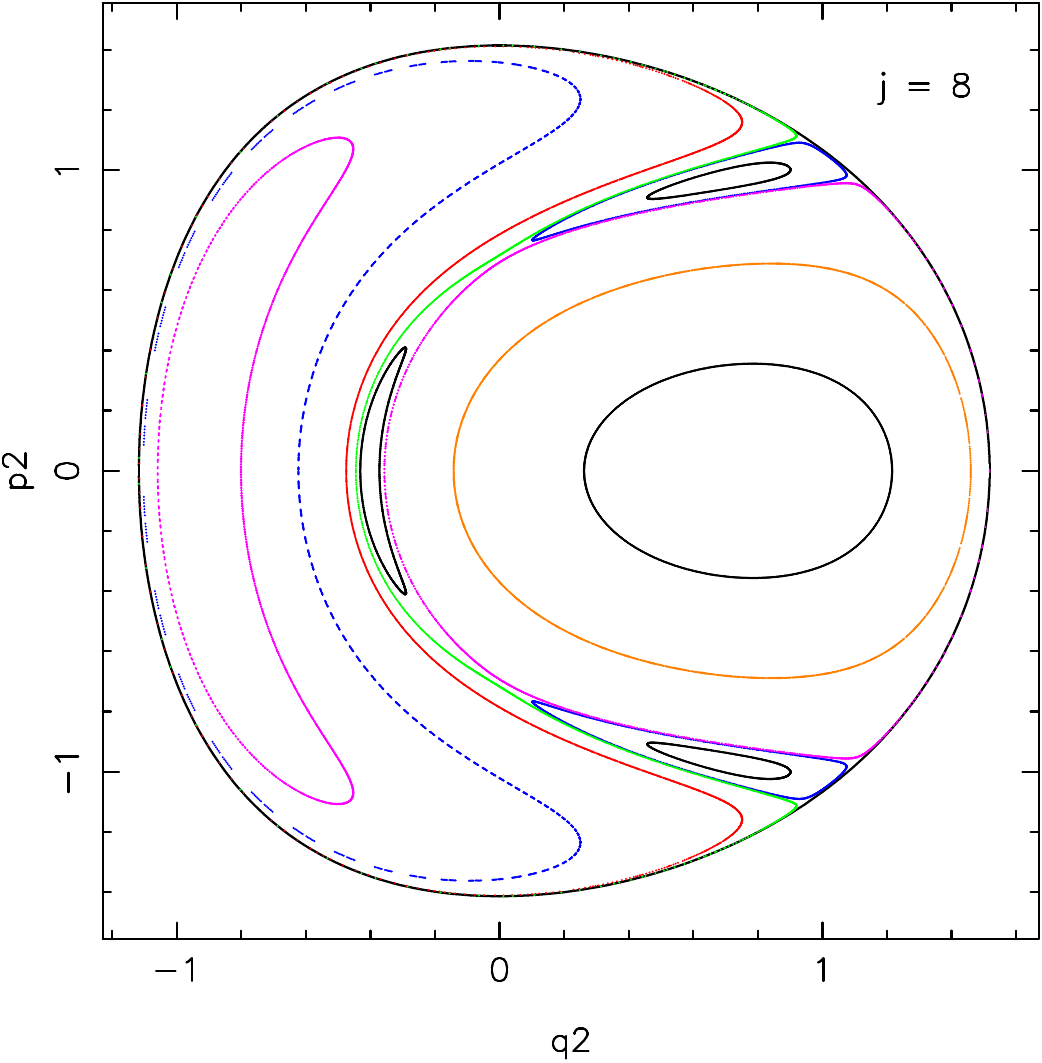}   }
  \caption{The Poincar\'e section of the model of three particles on a
    ring, with truncated Toda potential $\mathcal{T}_j$,
    $j=3,4,6,8$ (see the labels inside panels); energy $E=1$. }
\label{f4}
\end{figure}

\section{Numerical illustration}\label{sec3}

The purpose of this numerical section is to show quantitatively, and
visually, how quickly, by increasing $n$ at fixed $m$, the normalized
MLJ potentials $V_{nm}$ approach the Toda potential, and correspondingly, the
dynamics gets close to integrable. Comparison will be made with
the first polynomial approximations of Toda, denoted by 
$\mathcal{T}_j$, obtained by truncating the Taylor series of
$\mathcal{T}$ at order
$j$. Concerning $\lambda$, we shall use $\lambda=-2$
($\lambda$ should be negative, if we wish the steeper wall of the Toda
potential to stay on the left, as in MLJ potentials; 
$\lambda=-2$ corresponds to the quite common choice $\alpha=-1$ in FPU).

\vspace{3mm}\noindent
{\it A. Some coefficients.}\quad
Preliminarily, let us give a glance at the values of
the first few coefficients $k_{nm,j}$ 
entering the series expansion of $V_{nm}$. 
At small energy, the difference $V_{nm}-\mathcal{T}$ is dominated by
the fourth order term; correspondingly, the most
relevant quantity to look at is the
difference $\Delta_4=k_{nm,4}-1$. From \calleq{qjMie}, that we met in
the proof of the Proposition, it follows
\[
\Delta_4=\frac{2-nm}{(n+m+3)^2}\ .
\]
Computation shows that already for the classical values $m=6$ and
$n=12$, $\Delta_4$ is rather small, namely $\Delta_4\simeq -0.16$, and
raising $n/m$ to $4$ or $8$ lowers $\Delta_4$ to $-0.131$ or
$-0.088$.  Such values should be compared with the typical
constants used in FPU studies. In the standard FPU language, it is
\[
\Delta_4=\frac{\beta}{\beta_T}-1
=\frac{3\beta}{2\alpha^2}-1\ ;
\]
common values of $\Delta_4$, deduced from typical values of $\alpha$
and $\beta$ used in the literature, are much larger,\footnote{This is
  not surprising, since larger $\Delta_4$ accelerates the
  thermalization process and decreases the computational effort.}
namely $\Delta_4$ between $2$ and $6$ ($\Delta_4=-2/3$ in the original
FPU study, where $\beta=0$). Concerning the next coefficient
$k_{nm,5}$, a similar computation shows that the difference to $1$ is
$\Delta_5\simeq0.30$ already for $n=12$, while in typical FPU papers
the choice is $\gamma=0$, that is $\Delta_5=-1$.  We see that even for
not much large $n$, MLJ potentials are closer to
integrability than typical FPU models.  Figure \ref{f1} shows the behavior of
$\Delta_j$, $j=4,\ldots,7$, for $m=6$ and $n/m$ up to $32$.

\begin{figure}
  \centerline{\includegraphics[width=\figurewidth]{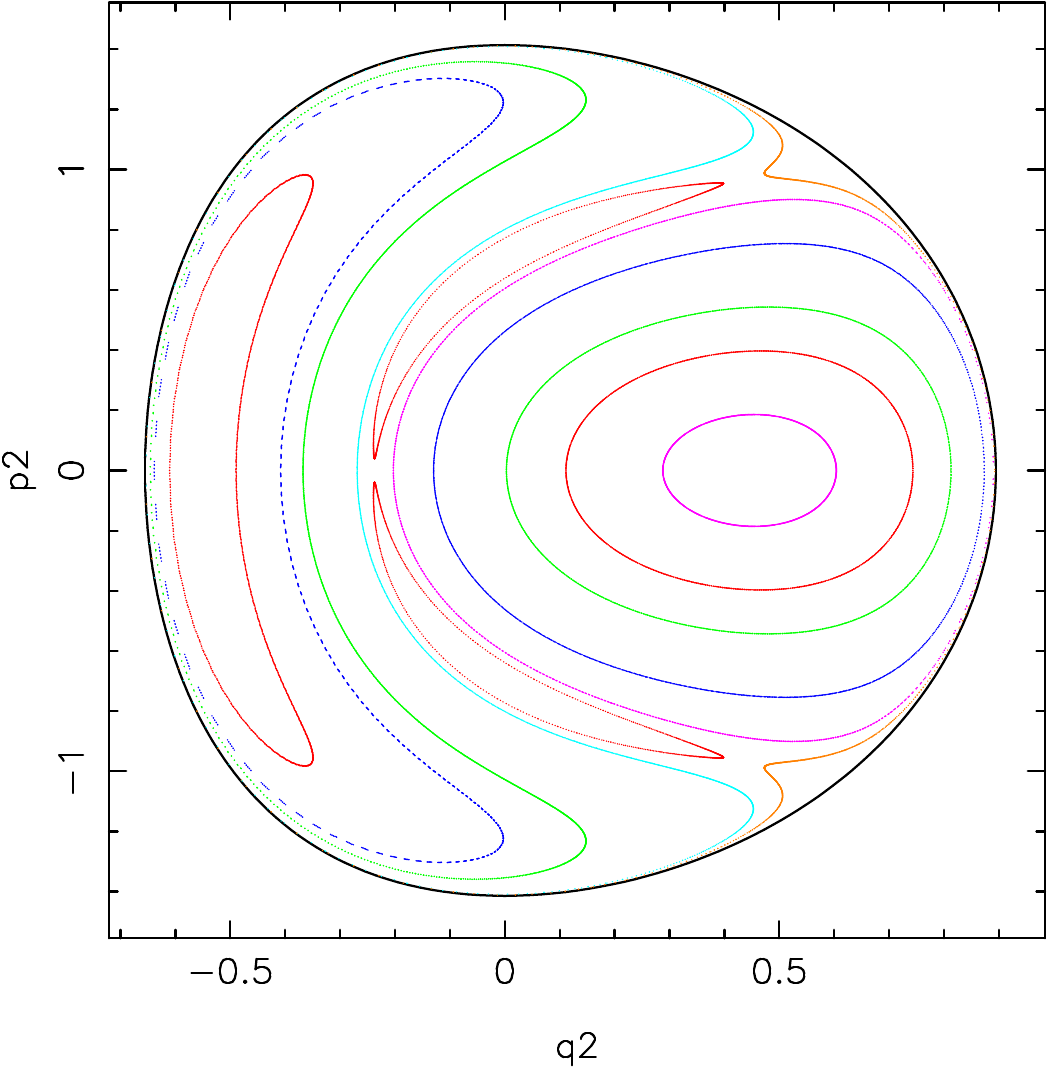}\hspace{1mm}
    \includegraphics[width=\figurewidth]{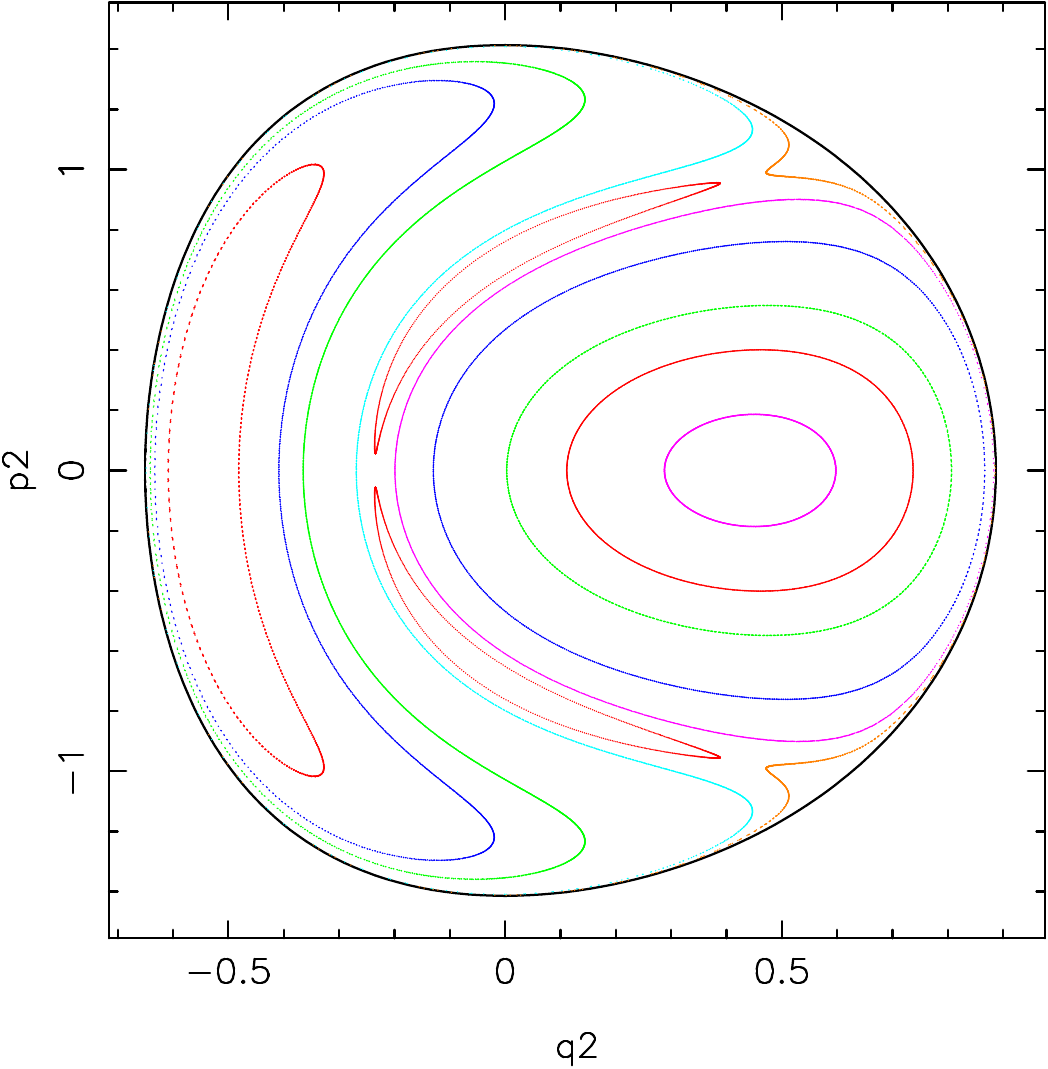}   }
  \caption{The Poincar\'e section of the model of three particles on a
    ring, with normalized MLJ potential $V_{nm}$; $m=6$ and $n=12$
    (left) and $48$ (right). Energy $E=1$.}
\label{f5}
\end{figure}

\begin{figure}
  \centerline{\includegraphics[width=\figurewidth]{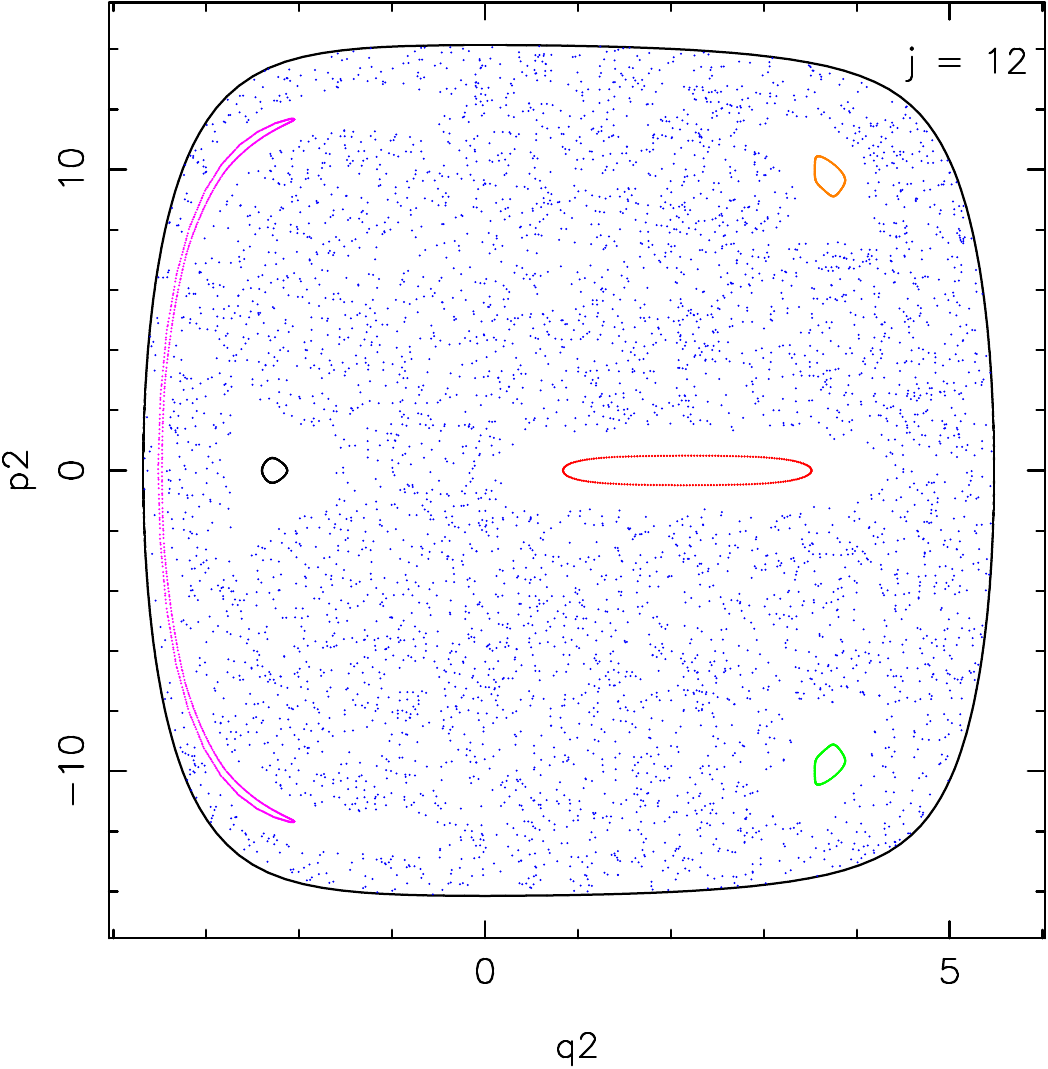}\hspace{1mm}
    \includegraphics[width=\figurewidth]{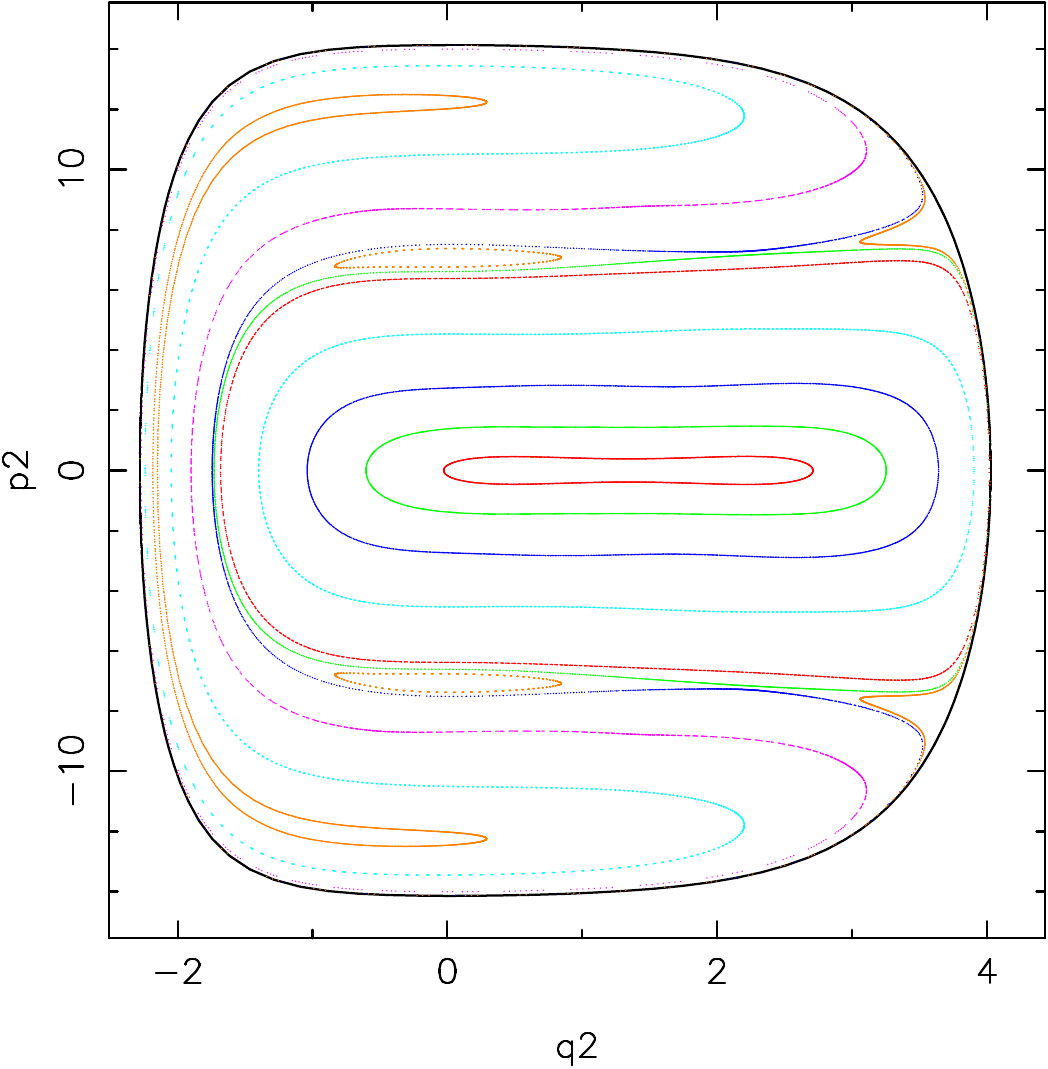}   }
  \caption{The Poincar\'e section of the model of three particles
    on a ring, at high energy $E=100$. Left: truncated Toda
    $\mathcal{T}_{12}$. Right: MLJ potential $V_{nm}$
    with $m=6$, $n=12$. }
\label{f6}
\end{figure}

\vspace{3mm}\noindent
{\it B. The shape of the normalized potential.}\quad
Figure \ref{f2}, left panel, shows how $V_{nm}$ (colored curves)
converges to $\mathcal{T}$ (black curve) by increasing $n$ at fixed
$m=6$; $n$ grows there from $12$ to $192$ ($2m$ to $32m$),
in geometric progression.
The right panel exhibits, on the same scale, Toda and its truncations
$\mathcal{T}_j$, $j=3,\ldots,10$.  The figure shows
that, for example, for potential energy around $1$,
even the common $12-6$ MLJ potential approximates Toda much better
than very exotic high order truncations $\mathcal{T}_j$. The
superiority becomes striking by growing energy. Let us stress that
high energies can localize in a single bond, even at small
specific energy, if the number of particles is large.

\vspace{3mm}\noindent
{\it C. Dynamics: three particles in a ring.}\quad
We come now to dynamics, and consider the model of three
particles on a ring:
\[
H(\zeta,\eta)=\sum_{\ind=0}^{2}\left[\frac{\eta_\ind^2}{2}
  +V(\zeta_{\ind+1}-\zeta_\ind)\right]\ ,
\]
with periodic boundary condition, baricenter at rest, two effective
degrees of freedom. This model was used in the celebrated paper
\cite{FST73}, with $V=\mathcal{T}$, to provide a strong numerical
indication that the Toda model is integrable. The method, as is typical
after \cite{HH} for systems with two degrees of freedom, consisted in analyzing
the Poincar\'e section. We do not provide details and refer to
\cite{FST73} for the choice of the Poincar\'e section and the coordinates
on it; everything is indeed absolutely standard.

Figure \ref{f3}, left panel, concerns the Toda model ($V=\mathcal{T}$)
and shows the Poincar\'e section for total energy $E=1$. The right
panel shows the same section for much higher energy $E=100$. The
absence of any chaotic region, no matter which is the value of $E$,
convinced the scientific community that the Toda model was integrable, and
prompted for mathematical proofs that soon arrived
\cite{M74,henon-toda,flaschka-toda}. Figure \ref{f4} refers to Taylor
truncations $\mathcal{T}_j$ of Toda, $j=3,4,6,8$ (see the labels inside
panels), at energy $E=1$. Let us recall that $\mathcal{T}_3$ coincides
with the celebrated H\'enon-Heiles Hamiltonian, up to a trivial
rescaling of energy by a factor $6$ (the first panel is indeed the
celebrated figure of ref. \cite{HH}, at $E=1/6$). Figure \ref{f5} refers
instead to the normalized MLJ potentials $V_{nm}$, for $m=6$
and $n=12$ (left), $n=48$ (right; very similar), at energy $E=1$. 
Not only, at this energy, the
chaotic region is absent, but the similarity with Toda, already for
$n=12$, is striking. At high energy, even high order truncations of Toda
behave completely differently from Toda; see figure \ref{f6}, left
panel, which represents the Poincar\'e section of $\mathcal{T}_{12}$
at $E=100$. Instead, at the same energy, $V_{nm}$ maintains an
excellent similarity with Toda even
for $m=6$ and $n=12$; see the right panel of the figure.

\begin{figure}
  \centerline{\includegraphics[width=\figurewidth]{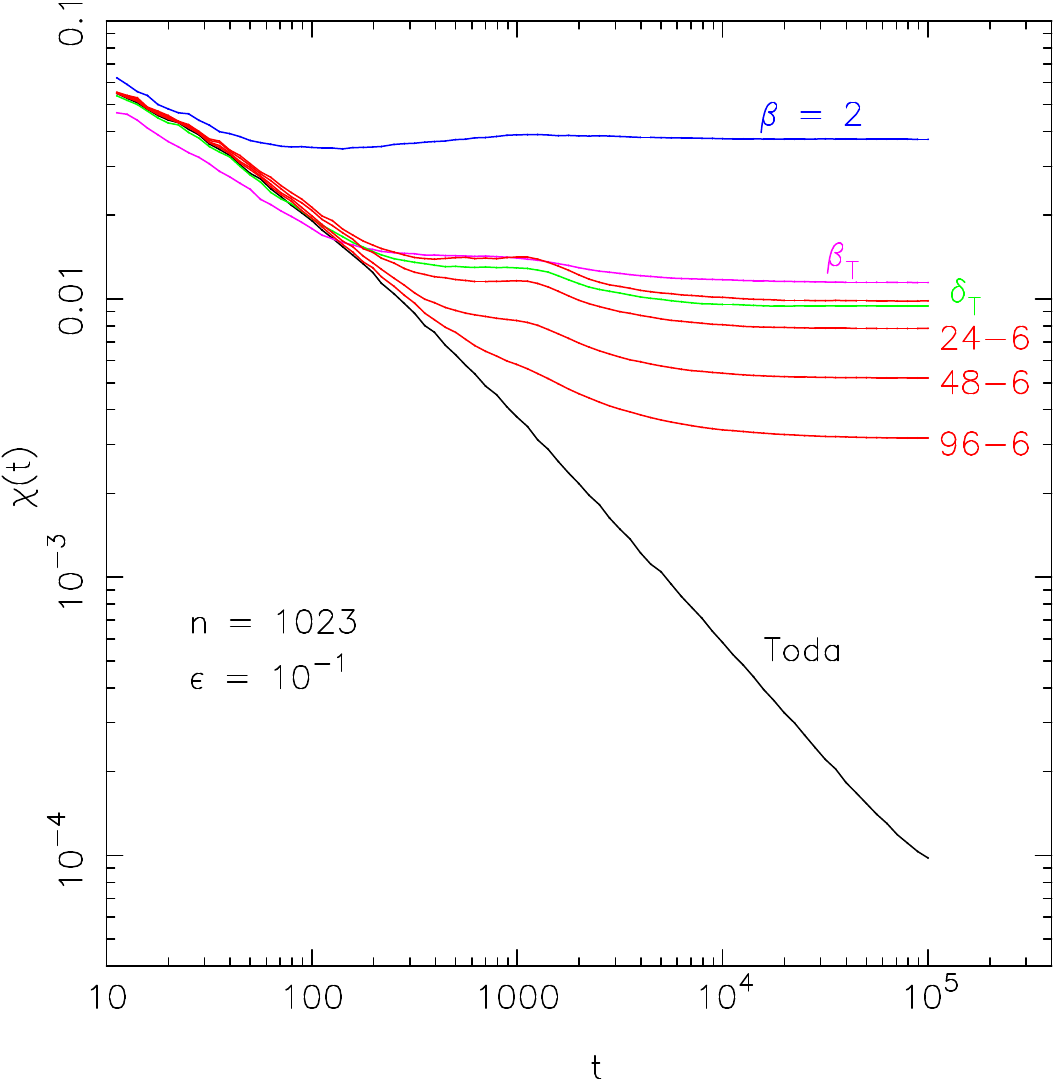}  }
\caption{The finite time Lyapunov indicator $\overline\chi(t)$
  vs. $t$, for Toda (black), FPU with $\alpha=-1$ and $\beta=2$
  (blue), truncations $\mathcal{T}_4$ and $\mathcal{T}_4$ of Toda
  (pink and green, respectively), then for MLJ potentials $V_{nm}$
  with $m=6$ and $n=12,\,24,\,48,\,96$ (red).}
\label{f7}
\end{figure}

\vspace{3mm}\noindent {\it D. Lyapunov exponents for large $N$.}\quad
Finally, we come to the dynamics for large $N$, actually $N=1024$. The
purpose is to quickly compare the dynamics of MLJ potentials $V_{nm}$,
and of Toda truncations $\mathcal{T}_j$, with the integrable Toda
dynamics, using as an easy tool the Lyapunov exponents (for a recent
extensive study of Lyapunov exponents in truncated Toda and other FPU
models, see \cite{BPP18,G23}).

Consider any initial datum $z$ in the phase space, let $F^t(z)$ be its
evolution at time $t$, and for any tangent vector $u$ in $z$, let
$DF_z^t u$ be the evolved tangent vector.
As is well known, the Lyapunov exponent $\chi(z,u)$ is defined as
\[
\chi(z,u)=\lim_{t\to\infty}\chi(t,z,u)\ , \qquad
\chi(t,z,u)=\frac1t \log \frac{\|DF_z^t u\|}{\|u\|}\ .
\]
For given $z$, essentially all vectors $u$ provide in the limit one
and the same value of $\chi(z,u)$, namely the maximal one. In fact,
very quickly the finite time quantity $\chi(t,z,u)$
loses the dependence on $u$, so we shall disregard it.  Unless there
is fully developed chaos, the dependence on $z$ is instead
effective. Experience however shows \cite{BPP18} that
taking an average even on a limited sample of points in phase space,
smooths significantly the $z$ dependence, and provides a reliable
finite time indicator $\overline\chi(t)$. We shall average on $24$
randomly chosen points, as in \cite{BPP18}.

Our aim here is not to perform a complete study, but just to
exemplify the theory, so we shall limit ourselves only 
to one value of the specific energy, namely $\varepsilon=0.1$. Figure \ref{f7}
shows $\overline\chi(t)$ vs. $t$ for Toda (black), for FPU with $\beta=2$
(blue) which has a contact of order $3$ with Toda,
for $\mathcal{T}_4$ (pink)
and $\mathcal{T}_6$ (green); then for MLJ potentials $V_{nm}$ with
$m=6$ and $n=12,\,24,\,48,\,96$ (red; $n=12$, almost coinciding with
$\mathcal{T}_6$, is not marked in the figure for lack of space).

For Toda, like for all integrable systems, $\overline\chi(t)$ goes to
zero as $\log t/t$. The other models reach instead a nonzero
limit. By increasing $n$, MLJ potentials $V_{nm}$ follow Toda for a
longer while. At this value of energy, even the $12-6$ MLJ approaches
Toda better than standard FPU, and similarly to the rather exotic
potential $\mathcal{T}_6$. For lower energies, however, the situation gets
complicated: the basic fact we aimed to illustrate, namely that by
increasing $n$ MLJ potentials approach better and better Toda, is
observed, and moreover, even for low $n$ they are closer to Toda than
standard FPU. But higher order truncations
$\mathcal{T}_j$, by lowering energy, become competitive, with a sort
of cross-over. We do not think it is worthwhile to further investigate
this point.

\medskip

\section*{Compliance with Ethical Standards}

\textbf{Competing interests} - The authors have no competing interests to declare that are relevant to the content of this article.

\noindent
\textbf{Data availability} - Data sharing is not applicable to this article as no datasets were generated or analysed during the current study.

\end{document}